\documentclass[prb,aps,twocolumn]{revtex4}

\usepackage{graphicx}

\begin{document}

\title{Linear scaling calculation of band edge states \\
  and doped semiconductors}

\author{H. J. Xiang}
\affiliation{Hefei National Laboratory for Physical Sciences at
  Microscale,
  University of Science and Technology of
  China, Hefei, Anhui 230026, People's Republic of China}

\author{Jinlong Yang}
\thanks{Corresponding author. E-mail: jlyang@ustc.edu.cn}

\affiliation{Hefei National Laboratory for Physical Sciences at
  Microscale,
  University of Science and Technology of
  China, Hefei, Anhui 230026, People's Republic of China}

\author{J. G. Hou}
\affiliation{Hefei National Laboratory for Physical Sciences at
  Microscale,
  University of Science and Technology of
  China, Hefei, Anhui 230026, People's Republic of China}

\author{Qingshi Zhu}
\affiliation{Hefei National Laboratory for Physical Sciences at
  Microscale,
  University of Science and Technology of
  China, Hefei, Anhui 230026, People's Republic of China}

\date{\today}

\begin{abstract}
  Linear scaling methods provide total energy, but
  no energy levels and canonical wavefuctions. From the density matrix 
  computed through the density matrix purification methods,
  we propose an order-N (O(N)) method for calculating 
  both the energies and wavefuctions of band edge states, which are
  important for optical properties and chemical reactions.  
  In addition, we also develop an O(N) algorithm to
  deal with doped semiconductors based on the O(N) method for band
  edge states calculation. We illustrate the O(N) behavior
  of the new method by applying it to  boron
  nitride (BN) nanotubes and BN nanotubes with an adsorbed
  hydrogen atom.
  The band gap of various BN nanotubes are investigated systematicly
  and the acceptor levels of BN nanotubes with an isolated adsorbed H
  atom are computed.
  Our methods are simple, robust, and especially suited for the application in
  self-consistent field electronic structure theory. 
\end{abstract}

\maketitle

\section{Introduction}
\label{intro}
Traditional electronic
structure algorithms calculate all eigenstates associated with
discrete energy levels. The disadvantage of this
approach is that it leads to a diagonalization problem
that has an unfavorable cubic scaling in the computational effort. 
Linear scaling density functional or Hartree-Fock methods
are an essential tool for the calculation of the electronic structure
of large systems containing many atoms.\cite{DM_rev} The key point of the success
of most linear scaling methods is that only the density
matrix or localized Wannier functions which span the occupied manifold is
calculated. In these O(N) methods, no canonical wavefunctions or
eigenvalues are available. However, in many cases, one may be
interested in some eigenstates, especially the states near the Fermi
level, i.e., band edge states. For instance, from the theory of
frontier orbitals, 
many molecular properties are determined by the highest
occupied molecular orbital (HOMO) and lowest unoccupied molecular
orbital (LUMO), and 
frontier orbitals play an important role in chemical reactions. 
On the other hand, there are some linear scaling algorithms such as
the Kim-Mauri-Galli (KMG)\cite{KMG} method need the Fermi level which can be
estimated from the HOMO and LUMO energy. 

There are some methods which can be used to
obtain band edge states. The most popular method for calculating states
near a reference energy $\epsilon_{ref}$ is the folded spectrum
method.\cite{Wang1994}  
However, in this method, by squaring the Hamiltonian, the condition 
number is also squared and thus the difficulty of solving the equation 
is also increased. To solve this problem, Tackett 
{\it et  al.}\cite{Tackett2002} 
presented the Jacobi-Davidson method in which the condition number and
difficulty in solving for the selected eigensolutions is the
same as the original eigenvalue equation. Unfortunately, the
implementation of the Jacobi-Davidson method is rather involved and
its application is not widespread. 
In the field of computational mathematics, the shift-and-invert
Lanczos algorithm is a well-known method for calculating a pair of eigenvalue
and eigenvector near a reference energy. This method was used by Liang 
{\it et  al.}\cite{Liang2003} to 
obtain the Fermi level in the context of linear scaling Fermi operator
expansion method.
In this method, the Lanczos
method is applied to the so-called shift-and-invert matrix, $(H-\epsilon_{ref}
I)^{-1}$, where $H$ and $I$ are the Hamiltonian and identity matrices,
respectively, and $\epsilon_{ref}$ is the reference energy. 
These matrices are not, of course, formed
explicitly. Instead, each time the Lanczos method requires a
multiplication of a vector $v$ by matrix $(H-\epsilon_{ref}
I)^{-1}$, a linear solver subroutine is called to solve the
corresponding
linear systems. If these linear systems are solved sufficiently
accurately, the convergence of the Lanczos method is typically much
faster compared to that when the matrix $H$ is used in the Lanczos
method. The difficulty now is that accurate numerical solution of
linear systems, needed on each iteration of the Lanczos method, can be
costly. Besides the difficulties of these methods mentioned above,  
when they are applied to get the frontier orbitals, another 
inconvenience is that a reference energy $\epsilon_{ref}$ must be
selected.       
 
Here in this work, we present an alternative simple method to get
states near gap based on linear scaling density matrix methods. In our
method, we do not need the reference energy. The new O(N) method is
particularly useful for calculating frontier orbitals in the framework 
of self-consistent field (SCF) electronic structure theory. Using this
method, we also propose a promising linear scaling method which can be
utilized to explore the energetics, defective levels, and gemoetry of
doped semiconductors.   

This paper is organized as follows: In Sec.~\ref{theory}, we
present our new O(N) methods for calculating band edge states and
dealing with doped semiconductors.
In Sec.~\ref{impl},
we describe the details of the implementation and perform
some test calculations to illustrate the rightness, robustness,
and linear-scaling behavior of our methods. 
In Sec.~\ref{appl}, we use our new methods to calculate the band gap of
boron nitride (BN) nanotubes and the acceptor level of a single H
adsorbed  BN nanotubes. Finally, our concluding remarks are given in
Sec.~\ref{con}.  

\section{Theory}
\label{theory}

\subsection{Calculation of band edge states}
Within our method, we must first obtain density matrix $\rho$
corresponding a given Hamiltonian $H$ before we proceed to calculate
band edge states. However, it is not an inconvenience in the framework
of linear scaling SCF electronic structure theory. 
In principles, any linear scaling density matrix 
methods can be used to obtain the density matrix.\cite{DM_rev,DM11,fd}   
Moreover, O(N) localized orbital based methods can also be used to
construct the density matrix.\cite{DM_rev,OM1,OM2}

In the representation of molecular canonical orbitals, 
density matrix $\rho$ and Hamiltonian $H$ are diagonal matrices of the
following forms:
\begin{equation}
  \begin{array}{lll}
  \rho &=& diag(1,1,\ldots,1,0,0,\ldots,0), \\
  H    &=&  diag(\epsilon_{1},\epsilon_{2},\ldots,\epsilon_{N_{e}/2},
  \epsilon_{N_{e}/2+1},\ldots,\epsilon_{N_{b}}), \\
  \end{array}
\end{equation}
where $N_{e}$ is the number of electrons of a closed-shell system,  
and $N_b$ is the number of basis functions. Without loss of
generality, we assume that
\begin{equation}
\epsilon_{1} \le \epsilon_{2} \le \ldots \le \epsilon_{N_{e}/2}
\le \epsilon_{N_{e}/2+1} \le \ldots \le  \epsilon_{N_{b}},
\end{equation}
then $\epsilon_{N_{e}/2}$ and $\epsilon_{N_{e}/2+1}$ will be the HOMO
and LUMO energies respectively.
It can be easily seen that:
\begin{equation}
  \rho H = H \rho =
  diag(\epsilon_{1},\epsilon_{2},\ldots,\epsilon_{N_{e}/2},0,\ldots,0).
\end{equation}
If $\epsilon_{N_{e}/2}>0$, then $\epsilon_{N_{e}/2}$ will be the
largest eigenvalue of $\rho H$. Otherwise, we can shift the Hamiltonian
H to $H+\lambda I$ ($\lambda>0$) so that $\lambda+\epsilon_{N_{e}/2}
>0$. Clearly, $\lambda+\epsilon_{N_{e}/2}$ is the largest eigenvalue
of $\rho (H+\lambda I)$. Using the similar argument, we can prove that 
if $-\lambda+\epsilon_{N_{e}/2 + 1} < 0$, $-\lambda+\epsilon_{N_{e}/2
  + 1}$ will be the smallest eigenvalue of $(I-\rho)(H-\lambda I)$.
We should note that the parameter $\lambda$ can be set to be a large
positive value without degrading the efficiency of the method. In
practice, we find that it is usually reliable by setting
$\lambda$ to be $1$ Ry. The largest (smallest) eigenvalue and
its corresponding eigenvector of  $\rho (H+\lambda
I)$ ($(I-\rho)(H-\lambda I)$) can be computed easily using the
well-known O(N) Lanczos method.
Up to now, we discuss the problem in 
the representation of molecular canonical orbitals $\psi$. In the
representation of orthogonal basis orbitals $\phi$, molecular
canonical orbitals $\psi$ can be expressed as
\begin{equation}
  \psi_i=\sum_\mu \phi_{\mu} C_{\mu i},
\end{equation}
where the coefficient matrix $C$ is a unitary matrix.
Thus in the
representation of orthogonal basis orbitals, density matrix $\rho_{or}$ and
Hamiltonian $H_{or}$ can be calculated as:  
\begin{equation}
  \begin{array}{lll}
  \rho_{or} &=& C \rho C^{+},\\
  H_{or} &=& C H C^{+}.\\
  \end{array}
\end{equation}
Moreover, $\rho_{or} H_{or}$ can also be obtained through a unitary
transformation of $\rho H$.
Since the unitary transformation of a matrix does not change its
eigenvalues, we can see that the above results deduced using the  
representation of molecular canonical orbitals also hold in
the representation of orthogonal basis orbitals. The procedure for
obtaining HOMO and LUMO states are illustrated in Fig.~\ref{fig1}(a).

Since many first principles codes use non-orthogonal atomic orbitals,
here we discuss the case of non-orthogonal basis.
A general method is transforming the non-orthogonal basis to orthogonal
basis. 
We achieve this by transforming the atomic orbital (AO) Hamiltonian matrix $H_{ao}$
to an orthonormal basis using $H_{or}=ZH_{ao}Z^{T}$ and obtaining the AO
density matrix $\rho_{ao}$ using $\rho_{ao}=Z^{T} \rho_{or} Z$, where the
inverse factor $Z=L^{-1}$, and $L$ is the Cholesky factor for which
$S=LL^{T}$. The Cholesky transformation has been used in severval 
linear scaling densit matrix programs.
We next show how to get wavefunction in the non-orthogonal basis. 
In non-orthogonal basis, a generalized eigenvalue problem should be
solved:
\begin{equation}
  H_{ao} \psi_{ao} = \epsilon S \psi_{ao},
\end{equation}
where $\psi_{ao}$ is the wavefunction in the non-orthogonal basis.
Given the wavefunction in the orthogonal basis $\psi_{or}$, which
satisfies 
\begin{equation}
  H_{or} \psi_{or} = Z H_{ao} Z^{T} \psi_{or} =  \epsilon \psi_{or},
\end{equation}
we have 
\begin{equation}
  \begin{array}{rll}
  H_{ao} Z^{T} \psi_{or} &=& \epsilon Z^{-1} Z^{-T} Z^{T} \psi_{or} \\
  &=& \epsilon  S  Z^{T} \psi_{or}, \\
  \psi_{ao} &=&  Z^{T} \psi_{or}.  
  \end{array}
\end{equation}

We also present another method to calculate band edge states in
non-orthogonal basis without transforming to orthogonal basis.
This method is particularly useful when localized orbitals 
based O(N) algorithms are employed. 
From $\rho_{ao} H_{ao} = Z^{T} \rho_{or}  H_{or} Z^{-T}$,
one can easily see that $\rho_{ao} H_{ao} \psi_{ao} = \epsilon \psi
_{ao}$ is equivalent to $\rho_{or}  H_{or} Z^{-T} \psi_{ao} =  \epsilon
Z^{-T} \psi_{ao}$. Thus $\rho_{ao} H_{ao}$ has the same eigenvalues as
$\rho_{or} H_{or}$. We can also prove that $\rho_{ao} (H_{ao} + \lambda S )$ has
the same eigenvalues as $\rho_{or} ( H_{or} + \lambda )$. 
Thus the largest eigenvalue of $\rho_{ao} (H_{ao} + \lambda S )$ will
be $\epsilon(\mathrm{HOMO})+ \lambda$.
We should
point out that $\rho_{or} ( H_{or} + \lambda )$ is hermitian, but
$\rho_{ao} (H_{ao} + \lambda S )$ is not. However, this doesn't pose
any problem since the Lanczos algorithm can also be used to get the extreme
eigenvalues of a non-hermitian matrix. We can see that the calculation
of HOMO state is simple since only $\rho_{ao}$, $H_{ao}$, and $S$ are
needed. However, the calculation of the LUMO state is a different
story. We can easily prove that $(I-\rho_{ao} S)(S^{-1}H_{ao}-\lambda I)$ has
the same eigenvalues as $(I-\rho_{or})(H_{or}-\lambda I)$ and
$-\lambda+\epsilon(\mathrm{LUMO})$ is  the smallest eigenvalue of $(I-\rho_{ao}
S)(S^{-1}H_{ao}-\lambda I)$. As can be seen, to calculate the LUMO
state, besides $\rho_{ao}$, $H_{ao}$, and $S$, we must also
have $S^{-1}$ or $S^{-1} H_{ao}$. The inverse of $S$ is usually a
formidable task. Fortunately, Gibson {\it et al.} introduced an O(N)
method to calculate $S^{-1} H_{ao}$.\cite{Gibson1993}

\subsection{Treatment of doped semiconductors}
To our best knowledge, most linear scaling methods are mainly applied to
semiconductors or insulators with an energy gap. When the system is
metallic or gapless, these O(N) methods fail or lose of effectiveness
since these methods rely on the sparsity of the density matrix and the
convergence of many of these methods is determined by the magnitude of
band gap. Partial occupation is another obstacle for many popular
linear scaling methods due to the non-idempotence of the density matrix.
Here we propose an O(N) method to deal with doped semiconductors where
dopants or defects exist. Our method has the similar spirit as that proposed by
Raczkowski and Fong in that a subspace larger than the occupied space
is used.\cite{Raczkowski2003} In their seminal work, the subspace optimization
method formulated in terms of localized nonorthogonal orbitals was
employed. However, besides two O(N$^3$) steps  in the Grassmann
conjugate gradient (GCG) algorithm, an additional O(N$^3$) step of
diagonalization is needed.
Another problem is that when the orbital localization is used to
acheive linear scaling, local minima might occur in the
subspace optimization method, resulting in a stalling of GCG
algorithm during the last several SCF steps.\cite{Raczkowski2003}

In our method, we treat the valence bands using the density matrix
method, and other defective bands are calculated using our O(N)
method for band edge states. For simplicity, we consider the cases
where only an electron or hole is present in a semiconductor, as shown
in Fig.~\ref{fig2}. In case of {\it n}-type
doping (Fig.~\ref{fig2}(a)), the total density matrix $\rho$ can be
calculated as   
\begin{equation}
  \rho = \rho_{val} + 0.5 | \psi_{N+1} \rangle \langle \psi_{N+1} |, 
\end{equation}
where $\rho_{val}$ is density matrix corresponding to the valence
band. In case of {\it p}-type doping (Fig.~\ref{fig2}(b)), the total density matrix $\rho$
can be calculated as
\begin{equation}
  \rho = \rho_{val} - 0.5 | \psi_{N} \rangle \langle \psi_{N} |.
\end{equation}
Both $\psi_{N+1}$ and $\psi_{N}$ are computed through the newly
developed O(N) method for band edge states.
Using the block Lanczos algorithm, our method can also be used when
several doped levels are present. In this case, the Fermi distribution  
can be used to get the occupation of doped levels.
Since the valence band are well
separated from the conduction band, $\rho_{val}$ is sparse, and  
the calculation of $\rho_{val}$ can be carried out using traditional
O(N) methods, such as the trace-correcting density matrix purification (TC2)
method.\cite{DM11} 
Since canonical orbitals $\psi_{N+1}$
and $\psi_{N}$ are usually delocalized, the total density
matrix $\rho$ is much denser than $\rho_{val}$. It is difficult to
deal with the full density matrix. However, we notice that in fact
only a small part of the full density matrix is used in the
construction of the new Hamiltonian. Thus in practice, we only
construct a small part of the full density matrix. To make our O(N)
method for the treatment of doped semiconductors
more clear, we show the flow-chart of a typical calculation in
Fig.~\ref{fig1}(b).  
Our method is very
simple and applicable to many doped systems. We should mention that    
our method is not a black-box method since some knowledge of the
studied system must be known prior. For instance, we should know the
doping type and number of doping levels. Typically, we can get this
information from intuition or deduction from other smaller systems
with similar characters.

\section{Implementation and Results}
\label{impl}
\subsection{Implementation}
Our newly developed method has been implemented in
SIESTA,\cite{siesta} a standard
Kohn-Sham density-functional program using norm-conserving
pseudopotentials and numerical atomic orbitals as basis sets.
In SIESTA, periodic boundary conditions are
employed to simulate both isolated and periodic systems.
Here we use the O(N) TC2
method\cite{DM11} to get the density matrix since it is very simple, robust, and
efficient.
The details about the implementation of the TC2 method
can be found in Ref. 13. 

In our O(N) method for doped semiconductors,
to obtain atomic forces, it is necessary to get the energy weighted
density matrix $E$ when using atomic basis sets. Take the case as shown in
Fig.~\ref{fig2}(a) as an example, 
\begin{equation}
  E= E_{val} +  0.5  \epsilon_{N+1}  | \psi_{N+1} \rangle \langle \psi_{N+1} |,
\end{equation}
where $E_{val}$ is calculated from $\rho_{val}$. For  energy weighted
density matrix $E$, we also compute and save only a part of the full
matrix. To speed up the calculation, we adopt the block Lanczos method
to calculate the defect levels, since the vectors produced by the previous
SCF step can be reused in the subsequent step. 
Usually, in the last several SCF steps, we don't need any matrix-vector
multiplications in the calculation of band edge states. Thus, 
when a geometry optimization is performed,
the extra amount for computing defect levels using our O(N) method is
almost negligible. This contrasts to the method proposed by
Raczkowski and Fong.\cite{OM1,OM2,Raczkowski2003}

\subsection{Validity and performance of the O(N) method for band edge
  states calculation}
All our calculations reported in this work are done in the
local density approximation (LDA).\cite{LDA} Unless otherwise stated,
the double-$\zeta$ plus polarization functions (DZP) basis set is used
in the calculations. 

We first validate our method by computing the HOMO and LUMO of H$_2$O
molecule. The energies of HOMO and LUMO are $-7.532$ ($-7.53257$) and
$-1.375$ ($-1.37292$) eV, respectively (values in parenthesis are
results from the diagonalization method). We also compare the HOMO and
LUMO wavefunctions with those from the diagonalization method, and
find that the agreement is remarkable.

To check the performance of our method, we calculate the HOMO and LUMO
of BN(5,5) nanotubes with different number of atoms in the
supercells. The CPU time used is shown in Fig.~\ref{fig3}. We can clearly
see the linear scaling behavior of our new method for both
single-$\zeta$ (SZ) and DZP
basis sets. 

For the purpose of comparison, we also calculate the LUMO of BN(5,5)
nanotube with 400 atoms using the folded spectrum method. The SZ basis
is adopted. Since the performance of the folded spectrum method
is very sensitive to the choice of the reference energy, several
different reference energies varying from the midgap position to 
the LUMO energy are chosen. The precision of the calculation is within
3 meV with respect to the value from the diagonalization. 
As shown in Fig.~\ref{fig4}, the CPU time used is very large,
especially when the reference energy is close to the LUMO energy (the
HOMO and LUMO energies are -7.075 and -2.577 eV respectively in the
current computing parameters setting).
Even when the the reference energy is chosen to be optimal, the
folded spectrum method is still slower by seventeen times than our new O(N)
method (387 s v.s. 22 s).

\subsection{Validity and performance of the O(N) method for doped
  semiconductors}
We will take BN(8,0) zig-zag nanotubes with an adsorbed H atom as an
example to illustrate the correctness and efficiency of our new
method. As shown by Wu {\it et al.}, a H atom prefers to adsorb on a B
atom, and the system is a $p$-type semiconductor.\cite{Wu2004} For a BN(8,0)
nanotube (128 atoms in the supercell) with an adsorbed H atom, the
energy difference between our result and that from the diagonalization
method is only 6 meV. And the force differences between our result and
that from the diagonalization method do not exceed 0.6 meV/\AA. 
Both the energy and forces agreement validates our new O(N) method
for doped semiconductors. We also deal successfully with a BN(8,0)
nanotube with two adsorbed H atoms on two B sites, indicating that our
method also works in case of systems with multi defect levels.

Here we show in Fig.~\ref{fig5} the CPU time
used in an ion step for supercells with different number of
atoms. Clearly, our new method for doped semiconductors displays a
linear scaling behavior. 

\section{applications}
\label{appl}
\subsection{Band gap of BN chiral nanotubes}
Previous study showed that for small zigzag (chiral angle $\alpha=0^\circ$) nanotubes the energy gap
decreases rapidly with the decrease of radius, while armchair
nanotubes (chiral angle $\alpha=30^\circ$) almost
have a constant energy gap. \cite{Xiang2003}
Although previous experiments\cite{zig-zag} indicated a preference for zig-zag and
near zig-zag BN tubes and a plausible explanation \cite{Xiang2003} was
proposed, a very recent high-resolution electron diffraction study on BN nanotubes
grown in a carbon-free chemical vapor deposition process
revealed a dispersion of the chiral angles.\cite{BN_chiral} Thus a
thorough knowledge of the dependence of the band gap upon the
chirality of BN nanotubes is desirable. Chiral BN nanotubes usually
contain large number of atoms in a unit cell, e.g., a BN(14,1)
nanotube has 844 atoms in the unit cell. These nanotubes are
difficult to be treated using traditional methods. Here we calculate
systematicly the band gap of BN nanotubes including chiral BN
nanotubes. Whenever the system is large enough to be sampled using
$\Gamma$-point, we use the new O(N) method for calculating band edge
states. The results are shown in Fig.~\ref{fig6}. Two
general trends are observed:
first for BN nanotubes with similar radius, BN nanotubes with larger
chiral angles have larger band gaps,  secondly, for BN nanotubes with chiral
angles close to zero, BN nanotubes with larger radius have larger band 
gaps. In addition, we can see that for BN($n$,$m$) nanotubes with
$n+m=k$, the band gap of BN($n$,$k-n$) does not depend monotonously
on the $n$ value due to the competition of the two trends
mentioned above, 
however, the band gap of BN($k-[\frac{k}{2}]$,$[\frac{k}{2}]$) (Here $[\frac{k}{2}]$
denotes the maximal integer no larger than $\frac{k}{2}$) nanotube is
the largest, and BN($k$,0) nanotube usually has the smallest band gap
except that the band gap of BN(8,2) nanotube is small than that of
BN(10,0) nanotube. The band gaps of some BN nanotubes were reported
previously and the results are in accord with ours,\cite{Xiang2003,Guo2005} 
and a more complete picture
for the trend of the band gap of BN nanotubes is presented here. 

\subsection{Acceptor level of H adsorbed BN nanotubes}
Wu {\it et al.}\cite{Wu2004} investigated the adsorption of a hydrogen atom on
zigzag BN(8,0) nanotube using a supercell containing 32 boron and 32
nitrogen atoms and found H prefers to adsorb on the boron atom which
introduces an acceptor state in the gap. 
They also showed that the dispersion of the
defect band is as large as 0.2 eV. Our test calculations in the
$\Gamma$-only approximation also show that the acceptor levels of a
single H adsorbed BN(8,0) nanotube using a 64-atoms or 128-atoms
supercell are 1.064 eV and 1.180 eV, respectively (Here, the acceptor
level is defined as the energy difference between the acceptor state
and the top of the valence band).  
In addition, the adsorption energy of the H atom also
depends on the chosen supercell: For instance, the adsorption energy
is $-0.353$ ($-0.246$) eV when using a 64-atoms (320-atoms) BN(8,0) supercell
and the diagonalization (our linear scaling) method.
All these facts suggest that larger supercells should be used to
predict the properties of BN nanotubes with an isolated adsorbed H atom.
Here with the O(N) method for doped semiconductors developed in this
paper, we can treat much larger radius BN nanotubes with truely isolated
adsorbed H atom through using huge supercells. Three BN nanotubes are
considered: BN(8,0) nanotube simulated using a supercell with 320
atoms, BN(15,0) nanotube simulated using a supercell with 720
atoms, and BN(13,2) nanotube with 796 atoms in the unit cell. 
Here we show 
the distribution of the acceptor state and the highest orbital of the 
valence band in Fig.~\ref{fig7}. Clearly, the acceptor state is a
relatively localized state around the adsorbed H atom, which agrees
with the result reported by Wu {\it et al.}\cite{Wu2004}. However, the
highest orbital of the valence band is delocalized and mainly
contributed by N 2p$_{\mathrm{z}}$ orbitals.
As can
be seen from Fig.~\ref{fig6}, 
BN(15,0) nanotube and BN(13,2) nanotube have similar radius but
different chirality, and the radius of BN(8,0) nanotube is smaller.  
The calculated acceptor levels introduced by an isolated H atom are
1.184 eV, 1.557 eV and 1.563 eV for BN(8,0), BN(15,0) and BN(13,2)
nanotubes, respectively. Thus the position of the defect level
is closer to the top of valence bands for smaller radius BN
nanotubes, but does not depend significantly on the chirality.

\section{Conclusions}
\label{con}
We present a simple O(N) method for calculating
band edge states using the density matrix obtained from O(N) 
electronic structure methods. 
Based on the O(N) method for calculating
band edge states, we further develop an O(N) algorithm to
deal with doped semiconductors.
In our methods, no reference energy is needed to
obtain the band edge states, and they are especially suited for the
application in SCF electronic structure theory. 
The O(N) behavior of the new methods is demonstrated by applying it to
bare and H adsorbed BN nanotubes. 
The band gap of various BN nanotubes are investigated systematicly
and the acceptor levels of BN nanotubes with an isolated adsorbed H
atom are calculated.
Our algorithms could be generalized straightforwardly to
spin-unrestricted systems,\cite{Xiang2005} such as magnetic
semiconductors and diluted magnetic semiconductors.\cite{spin}  

This work is partially supported by the National Natural Science Foundation of China
(50121202, 20533030, 10474087), by the USTC-HP HPC project, and by the
SCCAS and Shanghai Supercomputer Center.

\clearpage

\begin{figure}[!hbp]
  \caption{Schematic illustration of (a) the O(N) method for
    calculating the HOMO (the program flow for the LUMO calculation is
    similar except for some modifications as described in the text)
    and (b) the O(N) method for dealing with 
    doped semiconductors. Here ``density matrix'' 
    is abbreviated to ``DM''.}
  \label{fig1}
\end{figure}

\begin{figure}[!hbp]
  \caption{(Color online)Schematic illustration of the electronic
    structure of doped semiconductors: (a) $n$-typ doping and 
    (b) $p$-type doping.}
  \label{fig2}
\end{figure}

\begin{figure}[!hbp]
  \caption{Total CPU time for calculating HOMO and LUMO of BN(5,5) nanotubes
    using the linear scaling method.
    Here, both SZ and DZP basis sets are used. 
    All calculations were carried out on a 1.5 GHz
    Itanium 2 CPU workstation running RedHat Linux Advanced Server V2.1.}
  \label{fig3}
\end{figure}

\begin{figure}[!hbp]
  \caption{Total CPU time for calculating the LUMO of BN(5,5) nanotube
    with 400 atoms using the folded spectrum method with different
    reference energy. 
    Here, SZ basis set is used.}
  \label{fig4}
\end{figure}

\begin{figure}[!hbp]
  \caption{Total CPU time for calculating of BN(8,0) nanotube
    with a H atom adsorbed on a boron atom using the O(N) method for
    doped semiconductors.
    Here, double-$\zeta$ (DZ) basis set is used.}
  \label{fig5}
\end{figure}

\begin{figure}[!hbp]
  \caption{Band gap for various 
    BN($n$,$m$) nanotubes. BN($n$,$m$) nanotubes with $n+m=k$ are connected
    with a line to guide the eyes.}
  \label{fig6}
\end{figure}

\begin{figure}[!hbp]
  \caption{(Color online) (a) The acceptor state and (b) the highest
    orbital of the valence band of a BN(13,2) nanotube with an
    isolated adsorbed H atom. The insets show the enlarged plots
    around the  adsorbed H atom. }
  \label{fig7}
\end{figure}

\clearpage

\begin{figure}[!hbp]
  \includegraphics[width=6.5cm]{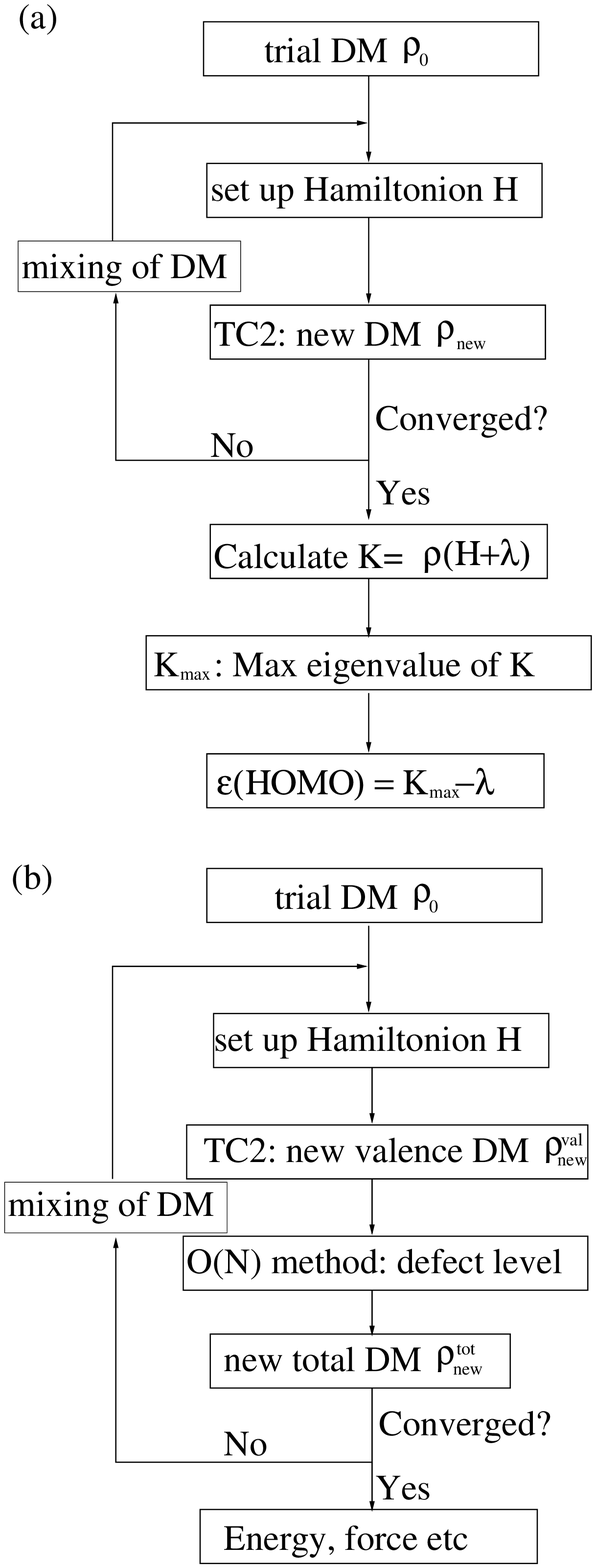}
\end{figure}
\begin{center}
  Fig. 1 of Xiang {\it et al.}
\end{center}

\clearpage
\begin{verbatim}







\end{verbatim}

\begin{figure}[!hbp]
  \includegraphics[width=6.5cm]{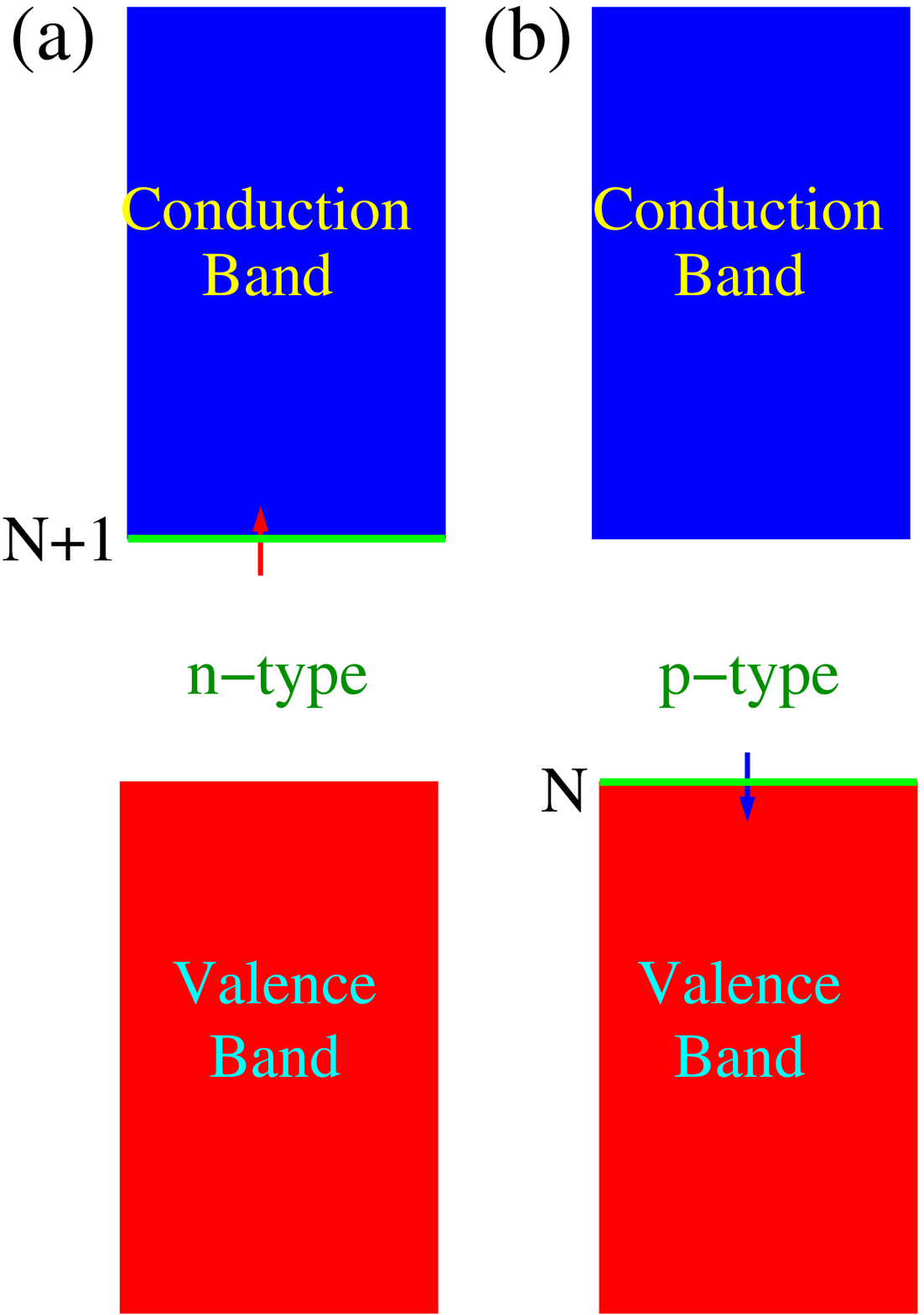}
\end{figure}
\begin{center}
  Fig. 2 of Xiang {\it et al.}
\end{center}

\clearpage
\begin{verbatim}









\end{verbatim}

\begin{figure}[!hbp]
  \includegraphics[width=8.0cm]{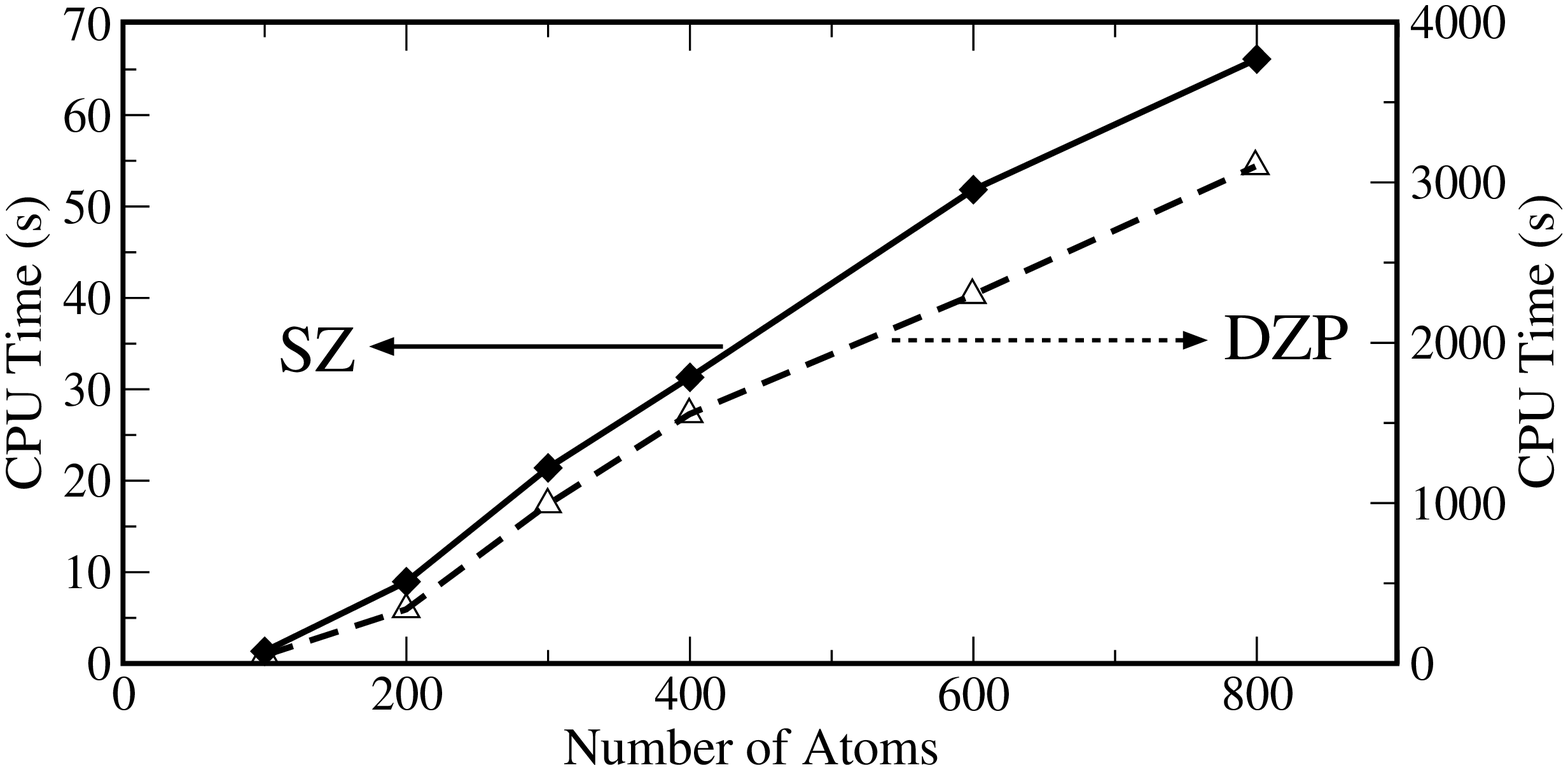}
\end{figure}
\begin{center}
  Fig. 3 of Xiang {\it et al.}
\end{center}

\clearpage
\begin{verbatim}









\end{verbatim}

\begin{figure}[!hbp]
  \includegraphics[width=8.0cm]{fig4.eps}
\end{figure}
\begin{center}
  Fig. 4 of Xiang {\it et al.}
\end{center}

\clearpage
\begin{verbatim}









\end{verbatim}

\begin{figure}[!hbp]
  \includegraphics[width=8.0cm]{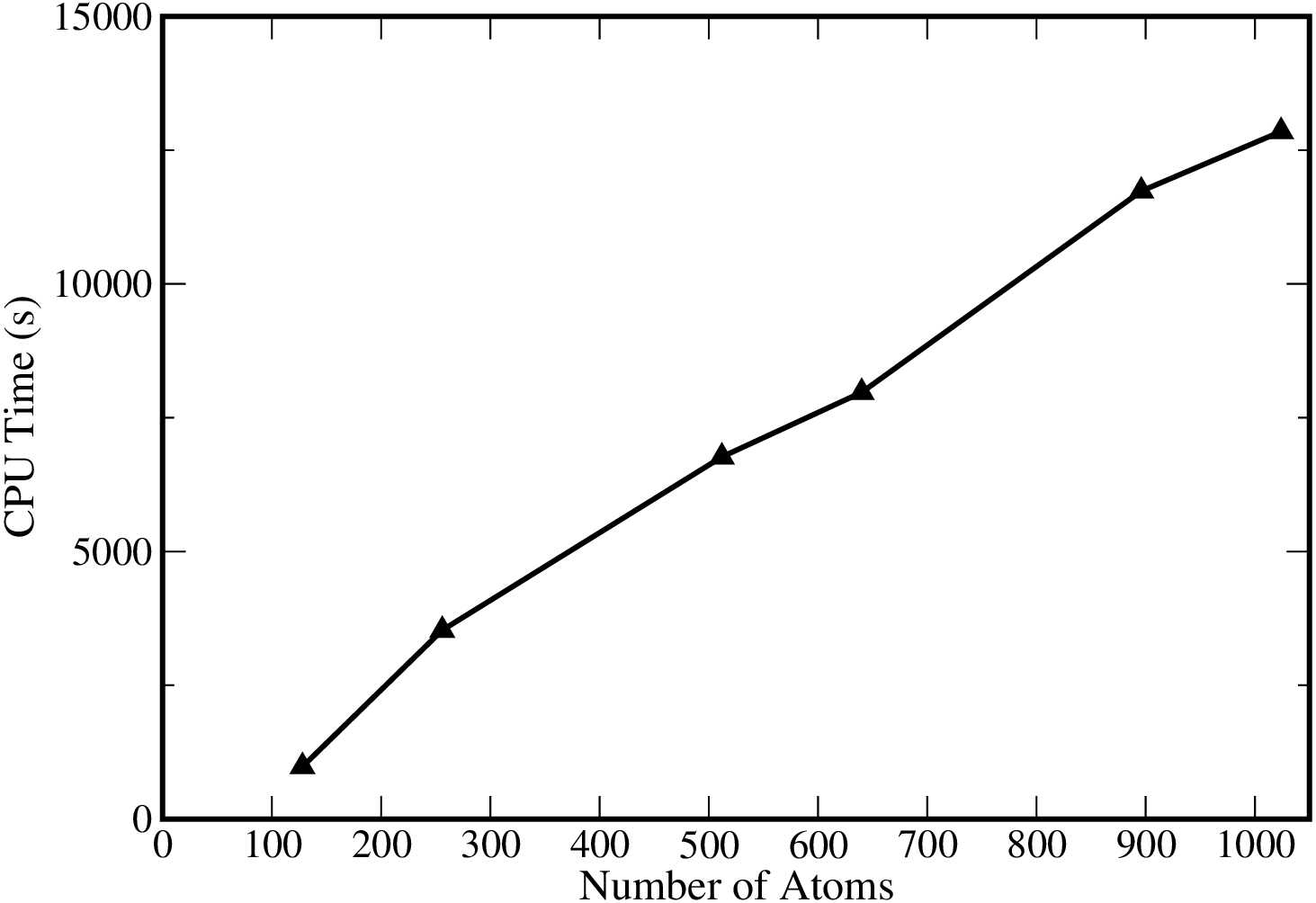}
\end{figure}
\begin{center}
  Fig. 5 of Xiang {\it et al.}
\end{center}

\clearpage

\begin{verbatim}









\end{verbatim}

\begin{figure}[!hbp]
  \includegraphics[width=8.0cm]{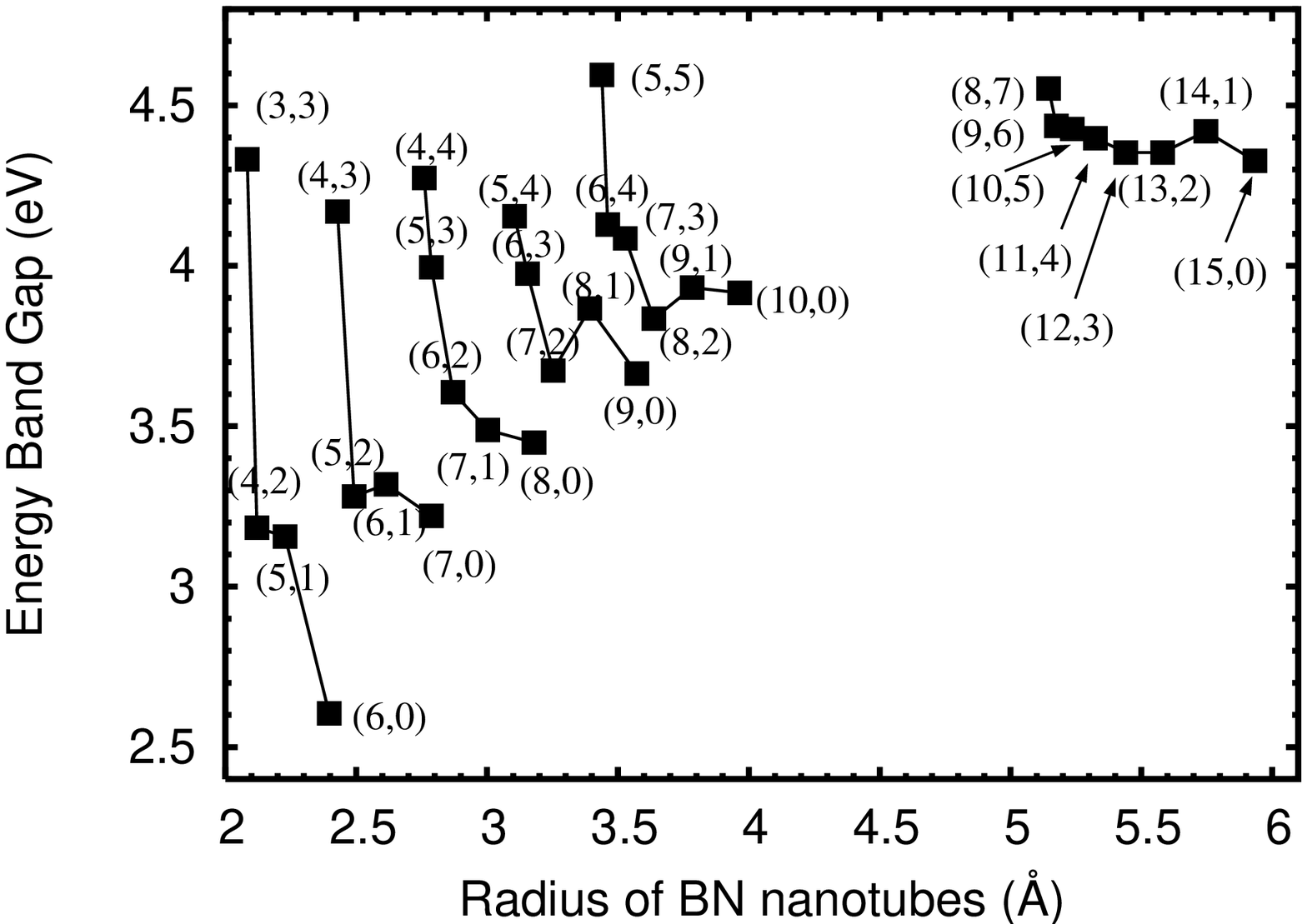}
\end{figure}
\begin{center}
  Fig. 6 of Xiang {\it et al.}
\end{center}

\clearpage

\begin{verbatim}









\end{verbatim}

\begin{figure}[!hbp]
  \includegraphics[width=8.0cm]{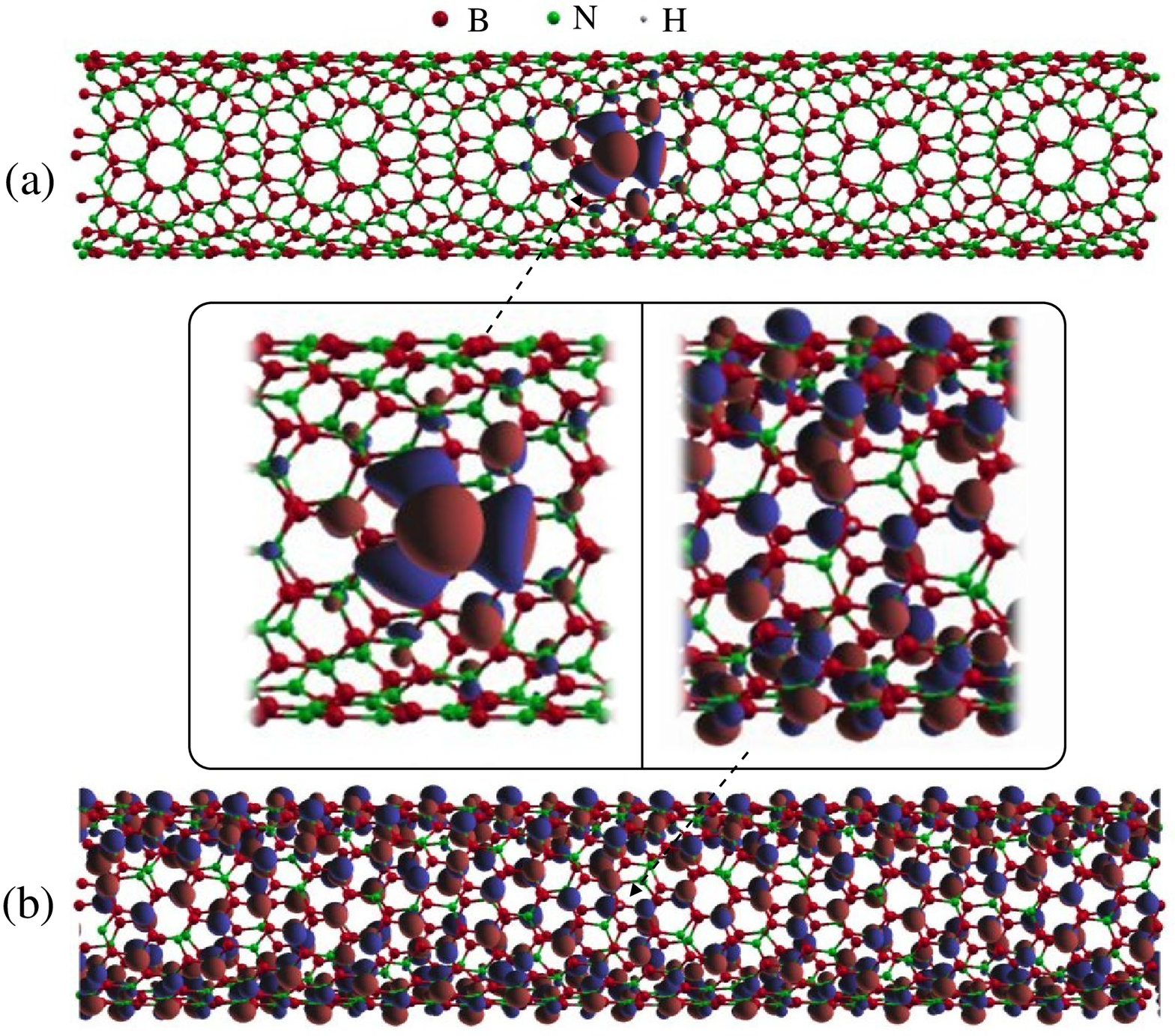}
\end{figure}
\begin{center}
  Fig. 7 of Xiang {\it et al.}
\end{center}


\begin{thebibliography}{99}

\bibitem{DM_rev}S. Goedecker, Rev. Mod. Phys. {\bf 71}, 1085 (1999);
S. Y. Wu and C. S. Jayanthi, Phys. Rep. {\bf 358}, 1  (2002).

\bibitem{KMG}J. Kim, F. Mauri, and G. Galli, Phys. Rev. B {\bf 52},
  1640 (1995).

\bibitem{Wang1994} L. W. Wang and A. Zunger, J. Chem. Phys. {\bf 100},
  2394 (1994).  

\bibitem{Tackett2002} A. R. Tackett and M. D. Ventra, Phys. Rev. B {\bf 66}, 245104 (2002).

\bibitem{Liang2003} W. Z. Liang, C. Saravanan, Y. H. Shao, R. Baer, A. T. Bell, and
 M. Head-Gordon, J. Chem. Phys. {\bf 119}, 4117 (2003).



\bibitem{DM11} A. M. N. Niklasson, Phys. Rev. B {\bf 66}, 155115
  (2002).

\bibitem{fd}J. R. Chelikowsky, N. Troullier, and Y. Saad,
  Phys. Rev. Lett. {\bf 72}, 1240 (1994).

  
\bibitem{OM1} F. Mauri and G. Galli, Phys. Rev. B {\bf 47}, 9973
  (1993); {\bf 50}, 4316  (1994).

\bibitem{OM2} P. Ordej\'on, D. A. Drabold, R. M. Martin, and
  M. P. Grumbach, Phys. Rev. B {\bf 51}, 1456(1995).


\bibitem{Gibson1993} A. Gibson, R. Haydock, and J. P. LaFemina, Phys. Rev. B {\bf 47}, 9229 (1993).

\bibitem{Raczkowski2003} D. Raczkowski and C. Y. Fong, Phys. Rev. B {\bf 68}, 014116 (2003).


\bibitem{siesta}J. M. Soler, E. Artacho, J. D. Gale, A. Garc\'{\i}a,
  J. Junquera, P. Ordej\'on, and  D. S\'anchez-Portal, J. Phys.:
  Condens. Matter {\bf 14}, 2745 (2002).

\bibitem{Xiang2005}H. J. Xiang, W. Z. Liang, J. L. Yang, J. G. Hou, and
  Q. S. Zhu, J. Chem. Phys. {\bf 123}, 124105 (2005).

\bibitem{LDA}D. M. Ceperley and B. J. Alder, Phys. Rev. Lett. {\bf
  45},
  566(1980); J. P. Perdew and A. Zunger, Phys. Rev. B {\bf 23}, 5048
  (1981).

\bibitem{Wu2004}X. J. Wu, J. L. Yang, J. G. Hou, and Q. S. Zhu,  
  Phys. Rev. B {\bf 69}, 153411 (2004);  J. Chem. Phys. {\bf 121},
  8481 (2004). 

\bibitem{Xiang2003}H. J. Xiang, J. L. Yang, J. G. Hou, and Q, S.
  Zhu, Phys. Rev. B {\bf 68}, 035427 (2003).


\bibitem{zig-zag}D. Golberg and Y. Bando, Appl. Phys. Lett. {\bf 79},
  415 (2001);
  M. Terauchi, M. Tanaka, K. Suzuki, A. Ogino, and K. Kimura,
  Chem. Phys. Lett. {\bf 324}, 359 (2000).

\bibitem{BN_chiral} A. Celik-Aktas, J. M. Zuo, J. F. Stubbins, C. Tang
  and Y. Bando,  Appl. Phys. Lett. {\bf 86}, 133110 (2005).

\bibitem{Guo2005}G. Y. Guo and J. C. Lin, Phys. Rev. B {\bf 71}, 165402 (2005).

\bibitem{spin}S. A. Wolf, D. D. Awschalom, R. A. Buhrman,
  J. M. Daughton, S. von Moln\'{a}r, M. L. Roukes, A. Y. Chtchelkanova,
  and D. M. Treger, Science {\bf 294}, 1488 (2001); I. \v{Z}uti\'{c},
  J. Fabian, and S. D. Sarma,
  Rev. Mod. Phys. {\bf 76}, 323 (2004).
 

\end{thebibliography}
\end{document}